# Design study of a "scintronic" crystal targeting tens of picoseconds time resolution for gamma ray imaging: the ClearMind detector

**D. Yvon[a], V. Sharyy [a], M. Follin[a], J-P Bard[a], D. Breton[b], J. Maalmi[b], C. Morel[c], E. Delagnes[a]**

[a] *Univ Paris-Saclay, CEA, Institut de recherche sur les lois Fondamentales de l'Univers, 91191, Gif-sur-Yvette, France*

[b] *Univ Paris-Saclay, CNRS, Laboratoire Irène Joliot Curie, 91898, Orsay, France*

[c] *Aix-Marseille Univ, CNRS/IN2P3, CPPM, Marseille, France*

*E-mail*: dominique.yvon@cea.fr

ABSTRACT:

We describe the concept of a new gamma ray "scintronic" detector targeting a time resolution of the order of 25 ps FWHM, with millimetric volume reconstruction and high detection efficiency. Its design consists of a monolithic large $PbWO_4$ scintillating crystal with an efficient photocathode directly deposited on it. With an index of refraction higher for the photocathode than for the crystal, this design negates the total reflection effect of optical photons at the crystal/photo-detector optical interface, and thus largely improves optical coupling between the crystal and the photodetector. This allows to detect efficiently the Cherenkov light produced by 511 keV photoelectric conversions in $PbWO_4$, and to optimize the detector time resolution. Furthermore, the low-yield, fast scintillation light produced additionally by $PbWO_4$ increases the detected photon statistics by a factor 10, thus fostering accurate (3 dimensional) localization of the gamma ray interaction within the crystal and providing a fair measurement of the deposited energy.

This paper lists the technological challenges that have to be overcome in order to build this "scintronic" detector. We show that all the key technologies have now been demonstrated and present results of a preliminary Monte Carlo simulation, which include an innovative event reconstruction algorithm to support the claimed performances of the detector.

# I. Context

Recently, the development of new types of ultra-fast compact photo-detectors has improved coincidence time resolution (CTR) of scintillation spectrometric chains below 100 ps FWHM. These detectors consist of fast, thin scintillators, typically LSO, LYSO, $LaBr_3$ or $CeBr_3$, optically coupled to SiPM matrices [1], [2], [3], [4], [5].

That is why a new, very ambitious CTR technological frontier appears at the edge of 10 ps FWHM [6], which would make it possible to foresee new ultra-fast gamma ray imaging applications. For example, in positron emission tomography (PET), an image could be acquired virtually without tomographic inversion with a 10 ps CTR time-of-flight (TOF)-PET camera.

Detector efficiency is limited by the stopping power of the scintillating material, which depends on its density, its thickness and its effective atomic number. Today, energy and spatial resolutions are limited by scintillation light yield and efficiency of light collection. Spatial resolution degrades additionally because of scattering and multiple reflections of optical photons in the crystal. Time resolution is limited by the shape of the scintillation light pulse, namely its rise and decay times, and the efficiency in collecting these scintillation photons. As such, the generation of a few dozen Cherenkov photons by photoelectric or Compton electrons is almost instantaneous compared to the production of scintillation light. The collection of both the Cherenkov photons and the scintillation photons is impacted by reflections on the crystal surfaces. In addition, time resolution of the detector is also limited by the uncertainty on depth-of-interaction (DOI) in thick crystals.

# II. Detector development for PET

Detector developments for PET are numerous and rich. For example, the reader can refer to [7] for a general overview. Recent efforts have often been motivated by the following goals:

- combining PET and magnetic resonance imaging (MRI) together,
- improving spatial resolution of PET (often in the context of neurological or of small animal imaging),
- improving time resolution of TOF-PET camera that permits to reduce patient dose or scan duration, thus opening new opportunities in the field of low count imaging as for theranostics [8].

The technologies under development must achieve a delicate compromise between all the properties described in the previous paragraph, especially the efficiency constraint, sometimes aiming at several of these goals.

Most of these developments use inorganic scintillators with high photon yields coupled to semiconductor photo-multipliers (SiPMs). SiPMs provide excellent time resolution, have a high quantum efficiency of photodetection and retain good properties when placed in an intense magnetic field: a major requirement for PET-MRI combined scanners. They are relatively simple to use. Their main drawback is their large dark count rate, which implies either to use scintillators with high light yields, or to cool them down a lot. SiPMs have become the reference photodetector for many manufacturers and the replacement of PMTs by SiPMs allowed to improve TOF-PET scanner CTRs from ~550 ps FWHM [9], [10] down to ~350 ps FWHM [11], [12]. Very recently, the Biograph Vision scanner from Siemens achieved a CTR of 214 ps FWHM [13].

## i. R&D on spatial resolution

R&D on spatial resolution represents a more limited effort. In nonTOF 3D PET, improving spatial resolution at constant image signal-to-noise ratio (SNR) requires to increase scan statistics by the fourth power of the spatial resolution improvement factor, *e.g.* by increasing geometrical acceptance of the scanner, which combines detection efficiency and detector angular coverage, or by increasing scan duration. As an example, the EXPLORER project proposed to extend the axial length of the scanner to 2 m to improve sensitivity [14]. It is a challenge that scintillator technologies are struggling to reach. Some breakthrough technologies using liquid detectors are proposed by the community of physicists. Pre-localization of the annihilation position by using TOF information [15] or three-photon imaging techniques [16] allows for reducing this constraint on scan statistics by a factor proportional to the ratio of the imaged object size to the pre-localization accuracy [17], [18].

### ii. R&D on time resolution

R&D on time resolution is now the main direction in research laboratories working on gamma ray detectors for PET. [19], and [6] present a complete review of the developments of detectors for TOF-PET. If resolutions of 100 ps FWHM are likely to be demonstrated within the coming years, improving CTR by an order of magnitude would need to introduce a paradigm shift in PET instrumentation [20]. The sensors currently used for PET are far from achieving such a performance. This is due in particular to "non-optimal" light production yield, "too slow" scintillation rise and fall time constants, and to the significant thickness of crystals required for efficient detection (typically 2 cm) of 511 keV annihilation photons, which induces a dispersion in the light collection times. To overcome these limitations, the community proposed to:

- improve scintillator light yield: a long R&D process,
- use Cherenkov light: a small number of photons but almost instantaneous,
- measure the depth-of-interaction (DOI) of the gamma ray within the crystal and decorrelate detection time from DOI.

The prototypes proposed by the laboratories use one or more of these options. For example, [21] and [22] have studied the possibility to use the Cherenkov light to improve time resolution of a "slow" BGO detectors. [23] has documented efficient 511 keV gamma ray detection using Cherenkov light in a liquid detector. [24] have studied the performance of a $PbWO_4$ cubic Cherenkov detector decorrelating DOI by Monte Carlo simulation, [25] published excellent time resolutions using MCP-PMT with lead-glass optical windows, unfortunately with low 511-keV detection efficiency.

[26] and [27, 28] have particularly studied the feasibility and relevance of time-of-flight detectors based on lead fluoride ($PbF_2$) crystals. A Cherenkov radiator was assembled with a micro-channel plate photo-multiplier tube (MCP-PMT) using an optical gel. [27] have grouped signals in 16 channels amplified by low noise amplifiers (mini-circuit, 40 dB) and have read out these signals with the 32-channels SAMPIC module. They measured a detection efficiency of 25 % for a crystal $53 \times 53 \times 10$ mm$^3$ and a time resolution of 150 ps FWHM in the center of the detector. To pursue this progress, it became essential to calibrate the response of the photomultiplier and its complete readout chain over the entire detection surface [28]. This was achieved by using a pulsed laser with a light beam duration shorter than the MCP-PMT transit-time spread.

The potential of a total-body PET camera using $PbF_2$ crystals was modeled with the GATE Monte Carlo simulation platform [29], [30], [31]. Results show that despite its modest efficiency, such a detector would offer an equivalent imaging quality, or even slightly better as compared to other existing cameras [32].

To get a significant improvement on time resolution, it is necessary either to use photomultipliers with better time resolutions (nothing exists better than MCP-PMT), or to improve significantly the collection efficiency of the Cherenkov light produced in $PbF_2$. This second idea led the development of the ClearMind detector. Actually, the best way to improve optical coupling between the crystal and the photodetector is to deposit a photocathode directly on the crystal. Nevertheless, $PbF_2$ is fragile and, in addition, degases lead compounds and fluorinated lead when heated to the temperatures necessary for the photocathode evaporation, which is not suitable.

Lead tungstate ($PbWO_4$) is a scintillating crystal that has been studied in detail for particle physics experiments [33], [34], [35], [36]. It has much better mechanical properties and degases considerably less than $PbF_2$ at high temperature. For these reasons, it is considered to be a good candidate to have a bialkaly photocathode deposited directly on its surface. Its Cherenkov light production is close to $PbF_2$ and it also generates a few scintillation photons (~300 photons/MeV) at short time scales (a few nanoseconds). This increased light output, together with the improved optical coupling, will make it possible to measure the energy deposited in the crystal and to reconstruct the position of the gamma ray interactions, including in particular DOI.

We have also studied the possibility to use SiPMs as photo-detectors for $PbWO_4$. As expected, the dark count rate of the SiPMs makes this task very difficult. The use of a SiPM to detect Cherenkov light requires either an external trigger, for example a coincidence trigger with a MCP-PMT signal, or the means to reduce dark count rate.

**Table 1:** Main properties and targeted performances of a ClearMind detector.

| Specifications | Target performance @ 511 keV |
|---|---|
| Spatial resolution | Down to few $mm^3$ |
| Time resolution | 25 ps (FWHM) |
| Detector efficiency | Highest possible: $PbWO_4$ has a 9 mm attenuation length with a photoelectric fraction of 43 % at 511 keV |
| Plug and play once developed | |

# III. ClearMind Detector Design

To meet the specifications given in Table 1, we *propose to develop a position-sensitive detector consisting of a monolithic scintillating crystal on which a high efficiency photocathode is deposited (Figure 1).*

This "scintronic" crystal, which combines *scintillation* and *photoelectron generation,* optimizes the transmission of scintillation and Cherenkov photons to the photocathode without any optical coupling media (*e.g.* optical grease). Such a device will avoid internal reflection of optical photons on the crystal/photocathode interface thanks to the high refractive index of the photocathode.

The "scintronic" crystal will be encapsulated within a micro-channel plate based multiplier tube (MCP-MT) to amplify the signal and optimize the transit time of the photoelectrons towards the detection anodes, thus minimizing the time resolution of the detection chain.

The elegance of our detector consists in:

- improving the efficiency of light collection in a scintillating crystal of high density and high effective atomic number by depositing a photocathode directly on the surface of the scintillating crystal,
- encapsulating this "scintronic" crystal with a MCP-MT,
- making use of the Cherenkov light emission to improve time resolution of the detector,
- utilizing the map of photoelectrons produced at the photocathode to reconstruct the properties of the gamma ray interaction by means of robust statistical estimators and multivariate analysis.

We propose (Figure 1) to use a scintillating crystal with low light production yield (~300 photons/MeV), but very fast (main decay time of a few nanoseconds). This will allow us to detect Cherenkov photons without being dazzled by the slower scintillation light.

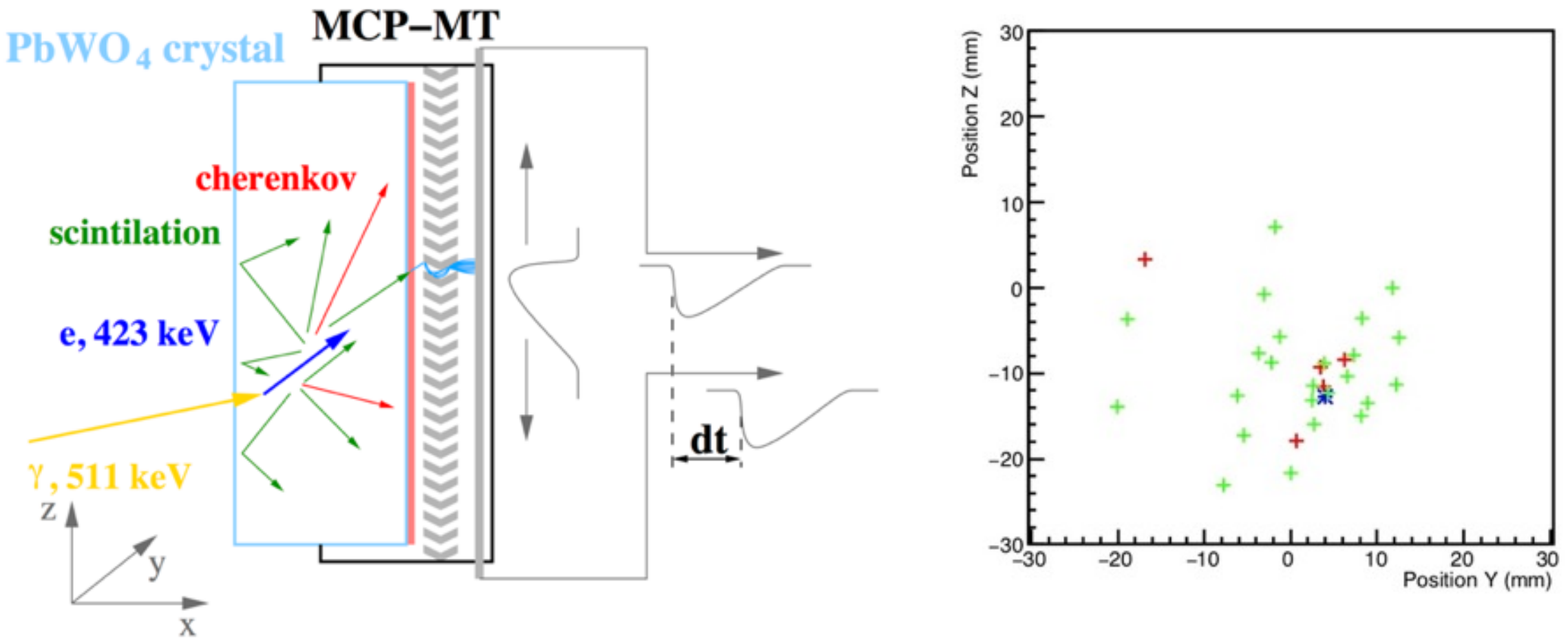

**Figure 1:** (Left) "Scintronic" crystal encapsulated within a MCP-MT. The photocathode is deposited directly on the scintillating crystal. The generated photoelectrons are amplified by a micro-channel plate. The amplified signals are collected on a densely pixelated anode plane read out by transmission lines and fast electronics [Kim, 2012]. (Right) Example of a photoelectron map produced on the photocathode by 511 keV gamma ray interactions within the crystal. The blue crosses mark energy deposits in the crystal, the red ones

correspond to the photons detected during the first 500 ps (mainly produced by Cherenkov effect) and the green ones mark scintillation photons detected between 500 ps and 20 ns

For this, we have measured the vapor pressures of the decomposition products of several crystals, "Cherenkov radiators" or scintillators, heated under vacuum (*i.e.* the process used for the photocathode deposition). As a result, we have selected lead tungstate ($PbWO_4$), which should allow to deposit the photocathode (*e.g.* bialkali, multialkali, etc.) directly on the scintillating crystal. The crystal would thus become the entrance window of the photomultiplier.

Depositing the photocathode directly on the scintillating crystal allows one to avoid the use of an optical coupling technology (*e.g.* optical gel) with a refractive index close to 1.5 and that of $PbWO_4$ is 2.3. This optical gel induces a strong loss of light by total reflection at the crystal/gel interface. The calculated total reflection angle value is 41°. This implies that assuming diffuse light is impinging at the optical interface, more than 75% of the photons lie in the solid angle where optical photons undergo total reflection. The photocathode refractive indices measured in the absorption band are close to 2.7. When the photocathode is deposited on the crystal, the transmission of photons from the crystal to the photocathode happens without total reflection. Hence, we expect a gain in the optical photon transmission probability by a factor up to 4.

The gain in optical coupling allows optimizing the time measurement based on Cherenkov light. The presence of a small number of scintillation photons associated with the use of a MCP and a densely pixelated anode plane makes it possible to acquire an image of the time-position map of photoelectrons produced on each instrumented face of the detector bloc (see Figure 1 right and Figure 7). This detector design has been patented [37].

### i. Photocathode on scintillating crystal

The first technological issue is the production of an efficient and stable photocathode on the surface of a $PbWO_4$ crystal. The reference in terms of efficiency to visible wavelengths (≥ 350 nm) remains the bialkali and multi-alkali photocathodes. However, these photocathodes are instantly deteriorated in contact of oxygen, even at very low concentrations. Therefore, the realization and the assembly of the components of the detector must take place under high vacuum.

### ii. Producing large size crystals

The second technological issue consists in *producing large-sized $PbWO_4$ scintillating crystals* up to $60 \times 60 \times 20$ mm$^3$. We have identified 3 companies producing $PbWO_4$ scintillation crystals: the Czech company CRYTUR (Turnov), and the Chinese companies SICCAS (Shangai) and EPIC (Jiangsu).

### iii. Signal readout

Finally, the third technological issue is related to the **number of pixels to read out on the anode plane of the MCP-MT**: a matrix of 1024 ($32 \times 32$) pixels or channels. To control the complexity, the heat dissipation and the electronics cost, we have connected these anodes to 32 transmission lines that are read out at both ends. The time difference in signal arrivals from both the edges allows to reconstruct the coordinate along the line and thus to divide the number of the channels by a factor 16. [38] have developed such a readout by using transmission lines to instrument high-yield scintillating crystals coupled with optical gel to micro-channel plate photomultiplier tubes (MCP-PMTs). They have shown that this configuration makes it possible to reconstruct the signals of the anode plane without measurable degradation of the MCP-PMT time resolution.

### iv. Photocathode deposition on $PbWO_4$ crystal

The deposition of a photocathode layer on a $PbWO_4$ crystal was carried out by the Bristish Photek company (St Leonards on Sea). After several tests, the work resulted in a great success: Photek has achieved a photocathode deposition on $PbWO_4$ with a quantum efficiency of 25 % at 400 nm.

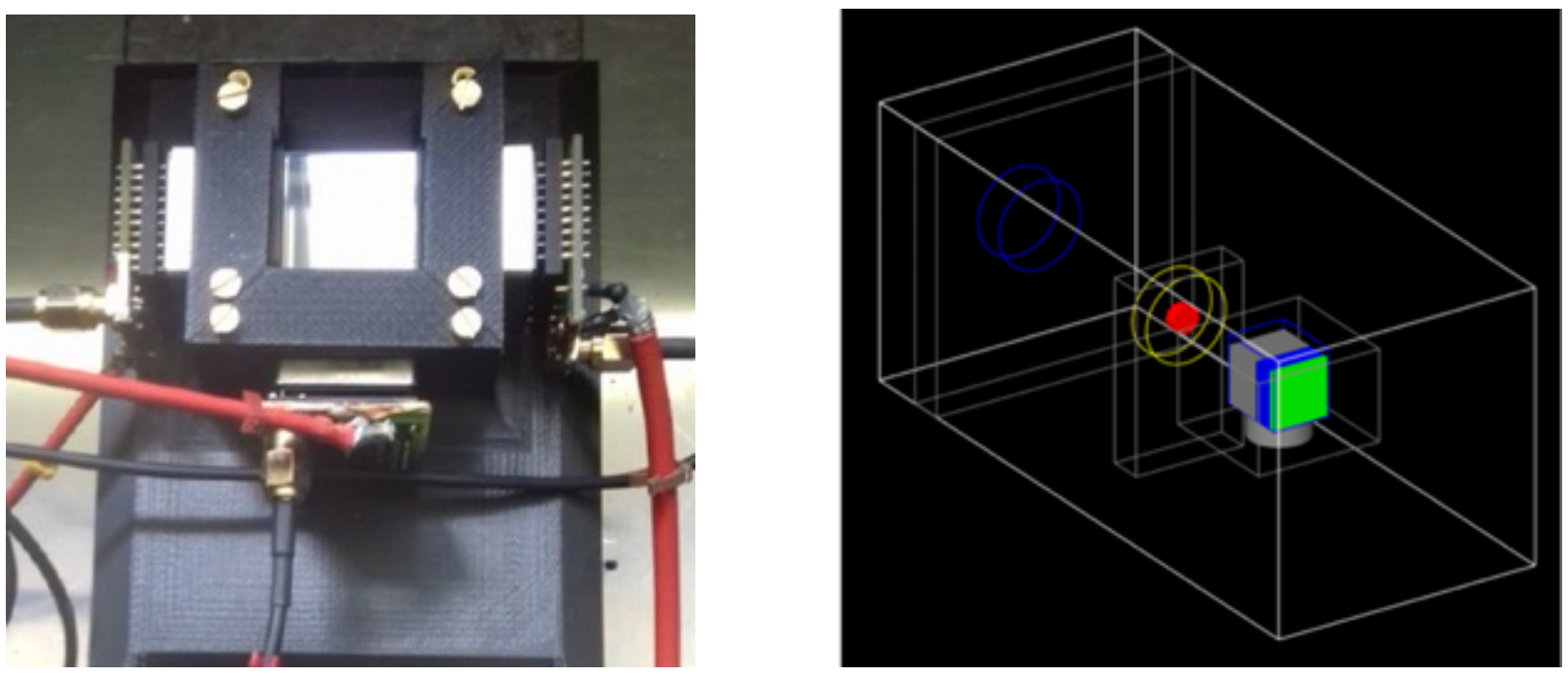

**Figure 2:** *Crystal scintillation properties measurements*. (Left) Picture of a cubic $PbWO_4$ crystal coupled with 3 Hamamatsu R11256-100 photomultipliers *using optical gel*. (Right) Simplified Monte Carlo simulation layout of the light yield measurement setup. We have used a 25 × 25 × 10 mm$^2$ $PbWO_4$ crystal grounded on all its faces, but the one coupled to the photomultiplier. We used a $^{22}$Na radioactive so urce and an additional YAP detector to tag the 511 keV photons from positron annihilations.

## v. $PbWO_4$ crystals light yield

We have measured and then compared scintillation light production yield and decay times of 3 $PbWO_4$ crystals with different doping from different suppliers (Table 2). The setup is shown at Figure 2. All the 3 crystals show quite similar performances consistent with our expectations. We are currently refining these measurements and preparing a dedicated paper documenting a temperature dependence study similar to the one performed for the Panda2 collaboration [36] and time constant measurements.

**Table 2:** Scintillation yield expressed as the ratio of the number of photoelectrons measured at room temperature with 3 $PbWO_4$ crystals to that calculated by Monte Carlo, assuming a production of 300 photons/MeV with a $PbWO_4$:Y emission spectrum (work in progress).

| Technology | Scintillation yield (%) |
|---|---|
| **Crytur: Panda2** | 0.97 |
| **SICCAS: CMS** | 0.81 |
| **SICCAS: Y-doped** | 0.87 |

## vi. MCP-PMT and transmission lines readout

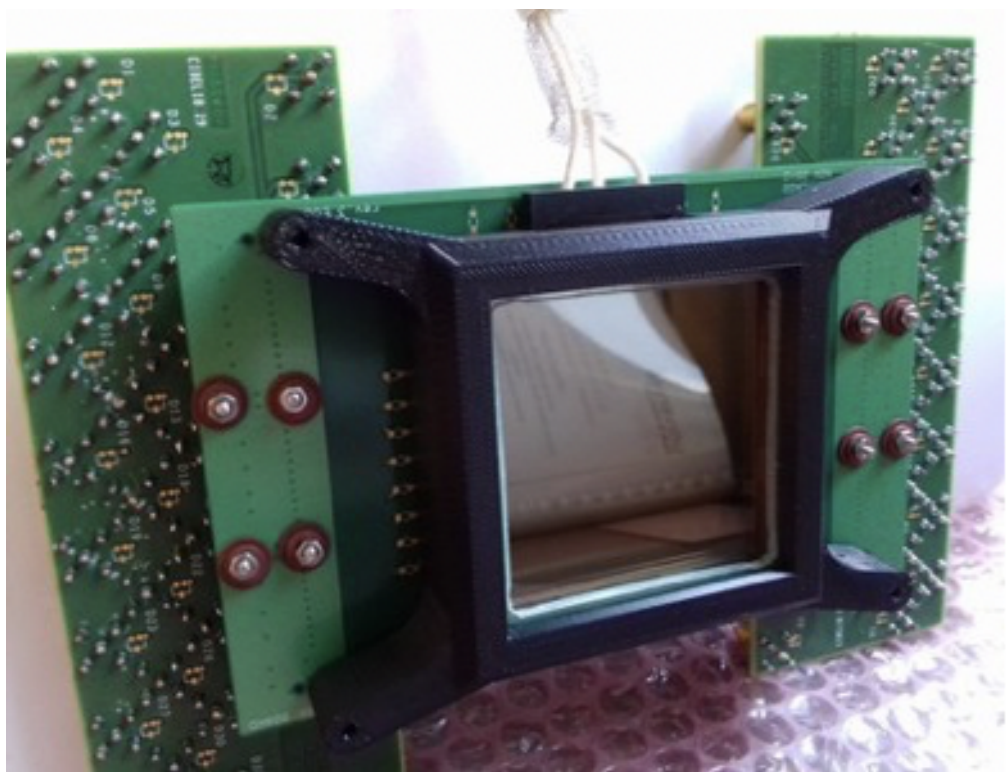

**Figure 3:** Picture of the MCP-PMT photodetector used for these developments.

A planacon XP-85122 MCP-PMT with 10 µm pore size and 1024 (32 × 32) anode pads was mounted on a transmission line readout board (Figure 3), provided by the university of chicago. The printed circuit board (PCB) was attached to the multi-pixelated anode pads of 1.1 × 1.1 mm$^2$ using a room-temperature bonded 3M$^{TM}$ conductive anisotropic tape (reference number 9705). Two additional boards were adapting the high density SAMTEC connector outputs to the 50 ohms SMA standard, which is more convenient for small scale prototyping.

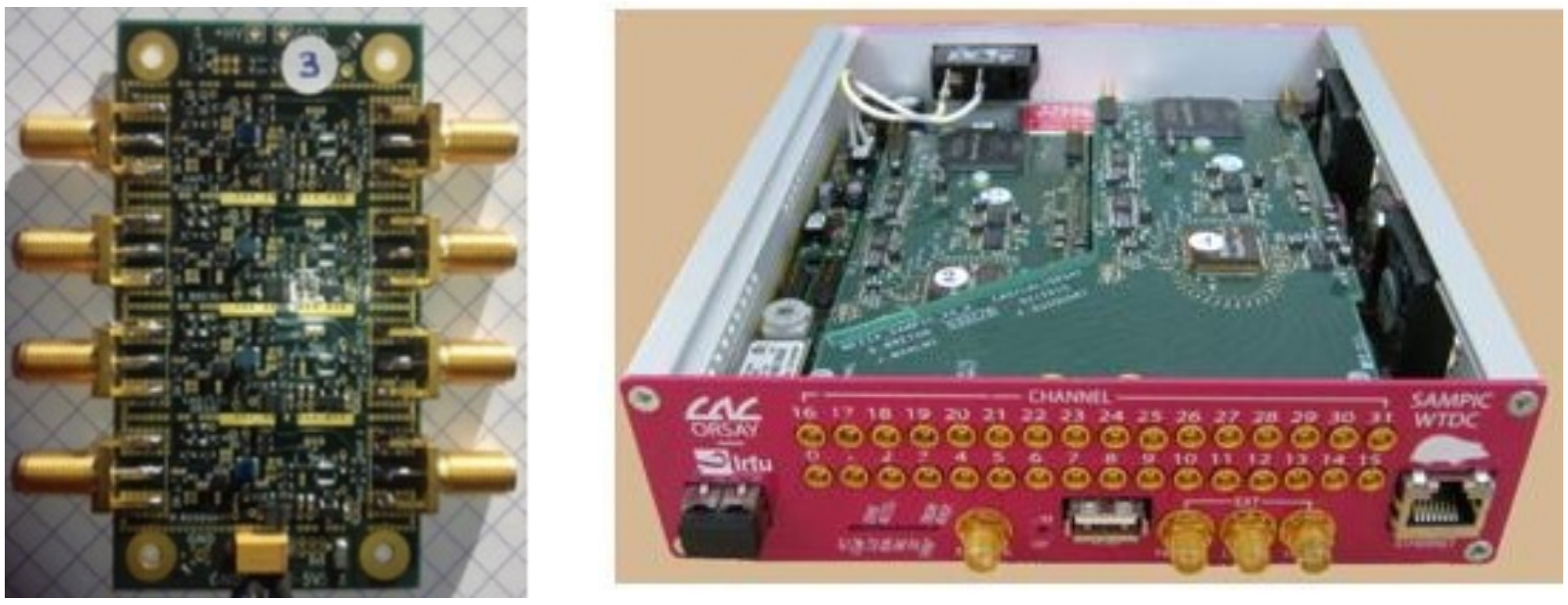

**Figure 4:** (Left) 4-channels fast amplifier prototype with a 700 MHz bandwidth and a gain of 100 used for the tests described in this work. (Right) 32-channels SAMPIC waveform recorder, which is currently used on many test benches all around the world [SAMPIC Workshop].

## vii. Fast readout amplifiers

We have developed prototypes of a fast multichannel amplifier board based on a broadband RFIC amplifier. The latter permits amplifying very fast negative pulses with a gain of 20 dB on 50 ohms with a moderate power supply (18 mA per channel). The measured bandwidth of 700 MHz offers a good compromise between noise and rise time for this detector. Special attention was paid to the channel powering and the PCB design in order to ensure a very low crosstalk between the channels. The 4-channel amplifier board shown in Figure 4 (left) houses two amplifying stages, hence presenting a gain of 40 dB.

## viii. SAMPIC waveform recorder

The 32-channels SAMPIC module (Figure 4, right) used in this setup is based on two 16-channels SAMPIC_V3C chips, themselves based on the new patented concept of waveform and time-to-digital converter (WTDC) [40], [41]. Each channel of the chip includes a DLL-based TDC to provide a rough time associated with an ultra-fast analog memory, which samples the signals between 1.6 and 8.5 GS/s. The recorded waveforms are used for precise timing measurements with a rms resolution of a few picoseconds. Every channel integrates a discriminator that can trigger it independently or participate to a more complex trigger. A first trigger level is implemented on-chip, while a second trigger level can be performed at the module level (32 channels). An associated data acquisition software is used to configure the module, start and stop acquisitions, store recorded data on disk in binary or ASCII format and visualize signal waveforms and parameter distributions. The SAMPIC module transfers raw signal waveforms to the computer, where all the necessary calibrations are applied on the fly by software. In addition, the latter performs one of the three hit time extraction algorithms: fixed threshold, constant fraction discriminator (CFD) or multiple CFDs [42]. In this work, we have run at 6.4 GS/s and used the CFD algorithm with a ratio of 0.5. Data stored on disk include signal waveforms (optional), calculated times and amplitudes, and selected SAMPIC parameters.

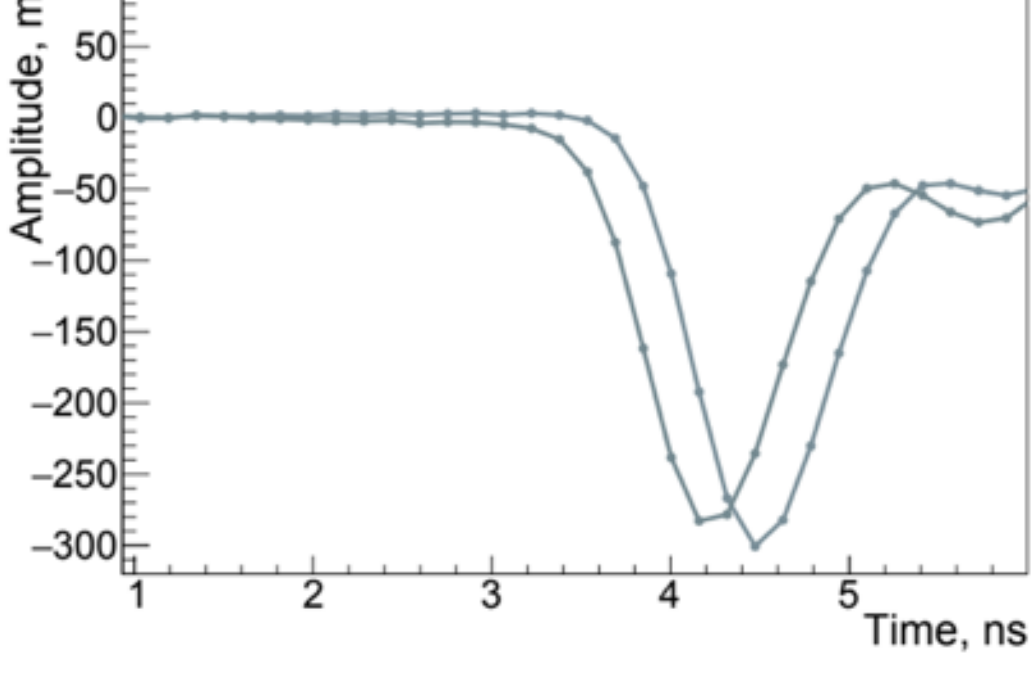

**Figure 5:** Single photoelectron as detected by the XP-85122 MCP-PMT. The pulses are registered by the SAMPIC Waveform recorder at both ends of a readout transmission line. Before the pulse, the signal shape is very flat, showing that the readout noise is minimal.

### ix. Photodetector performance

For a first test of this readout paradigm, we have acquired data from eight transmission lines. All the other transmission lines were grounded through 50 ohms at both ends. The MCP-PMT was mounted on a X-LRT-C computer-driven micrometric translation table from Zaber Technologies (Vancouver, Canada). A picosecond laser of type Pilas manufactured by the company NKT Photonics (Cologne, Germany) configured to generate single photoelectrons was lightening the MCP-PMT through a 300 µm pinhole placed at 2 mm of the optical window. This system allowed us to precisely position and move the MCP-PMT behind the pinhole.

We have measured very accurately the time differences between the two signals at the ends of the readout line for a pinhole centered on a readout line (Figure 5). The time difference distribution is gaussian and the measured widths range from 16 to 20 ps FWHM, depending on the readout line measured, but close to the SAMPIC module time resolution limit. From the measured signal speed along the readout line, we have determined spatial resolutions along the line that amount to 1.2 to 1.5 mm FWHM (Figure 6, left).

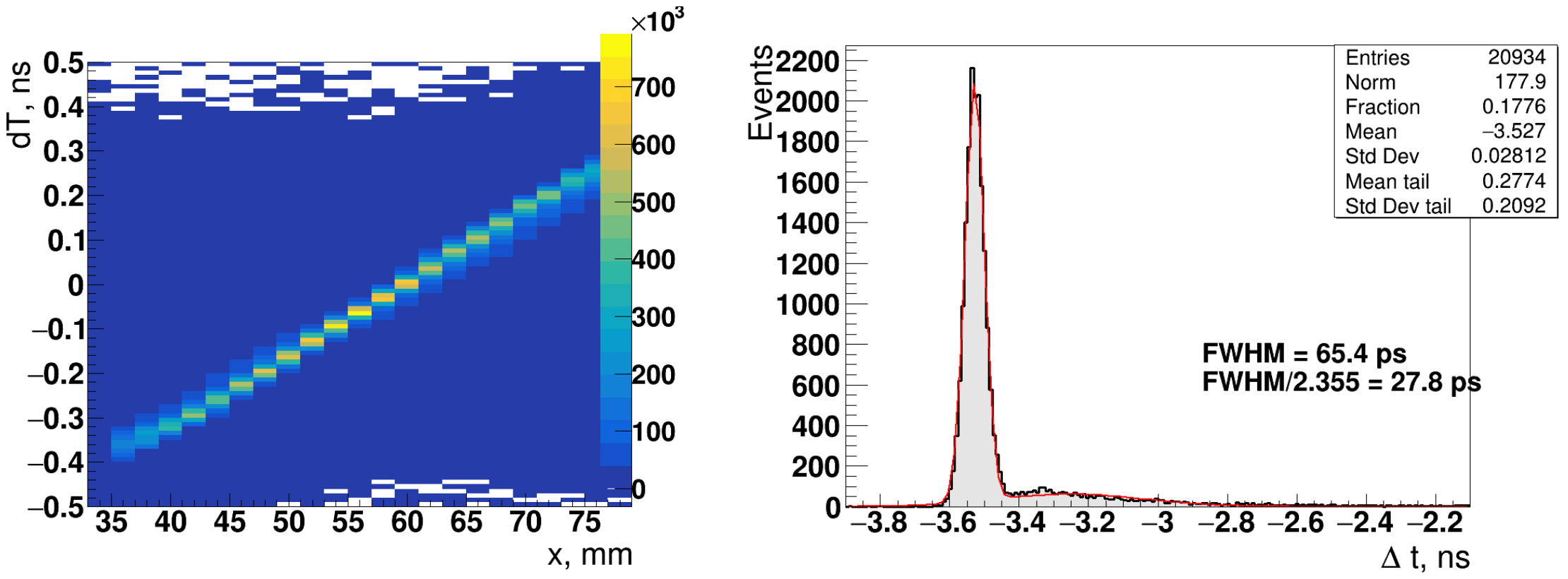

**Figure 6:** (Left) Histogram of position of the pinhole along the line versus time difference between signals registered at both ends of the targeted transmission line. (Right) Histogram of the time difference between the laser pulse and the average of the pulse times measured at both ends of the transmission lines.

The average of the pulse times measured at both ends of the transmission lines provides an accurate measurement of photoelectron detection time. The histogram of the time difference between the readout lines and the laser pulse, (Figure 6, right) shows a narrow main peak with a width of 67 ps FWHM and a long tail, typical of a photoelectron back-scattered at the entrance of MCP-PMT, which corresponds to 25 % of the total statistics.

## IV. Simplified detector Monte Carlo simulation

In order to optimize detector design and prepare future data analysis, we have written a simplified detector Monte Carlo simulation, which will be updated and refined as the hardware will evolve. We have used the CERN GEANT4 library [43] interfaced through the GATE macro language and libraries [29], [30]. We have assumed that the scintillating crystal will consist of a monolithic $PbWO_4$ crystal 60 × 60 × 10 (or 20) $mm^3$. Crystal properties were extracted from the literature, choosing $PbWO_4$:Y as a reference for this study. The refraction index was simulated according to [44]. We have averaged the ordinary and extraordinary refraction index in order to code mean optical propagation properties. The scintillation wavelength distribution was extracted from [45]. The attenuation length in $PbWO_4$ as function of the scintillation wavelength was modelled according to [35]. The $PbWO_4$ light yield was assumed to be 300 photons/MeV from data published in [46] and [33].

The two 60 × 60 mm$^2$ opposites faces of the crystal are covered by a blue-bialkali photocathode directly deposited on the crystal surface (first detector prototype will have only one). All the other crystal faces are assumed to be polished and black. The photocathode model, namely its refraction index, absorption length, and photoelectron production and extraction efficiencies were taken from [Motta, 2005].

511 keV gamma rays impinge on the crystal surface (Y and Z axis) and propagate along the X axis. Simulation of gamma ray interactions uses the Livermore library. Optical photons were produced by both the Cherenkov effect and the $PbWO_4$ scintillation.

GATE was configured to save data in the GEANT4/ROOT HitsFile format. A dedicated C++ software was then written to gather output data and store the time, 3D coordinates and energy *of every gamma ray interaction within the crystal*. Then, the absorption of optical photons (position, time and deposited energy) in the crystal and at the photocathodes were registered. This software also models the behavior of the photocathode according to [47] and computes the foreseen photoelectron production (time and 2D coordinates). This preliminary simulation did not include a model neither for the MCP-PMT photodetector, nor for the readout electronics.

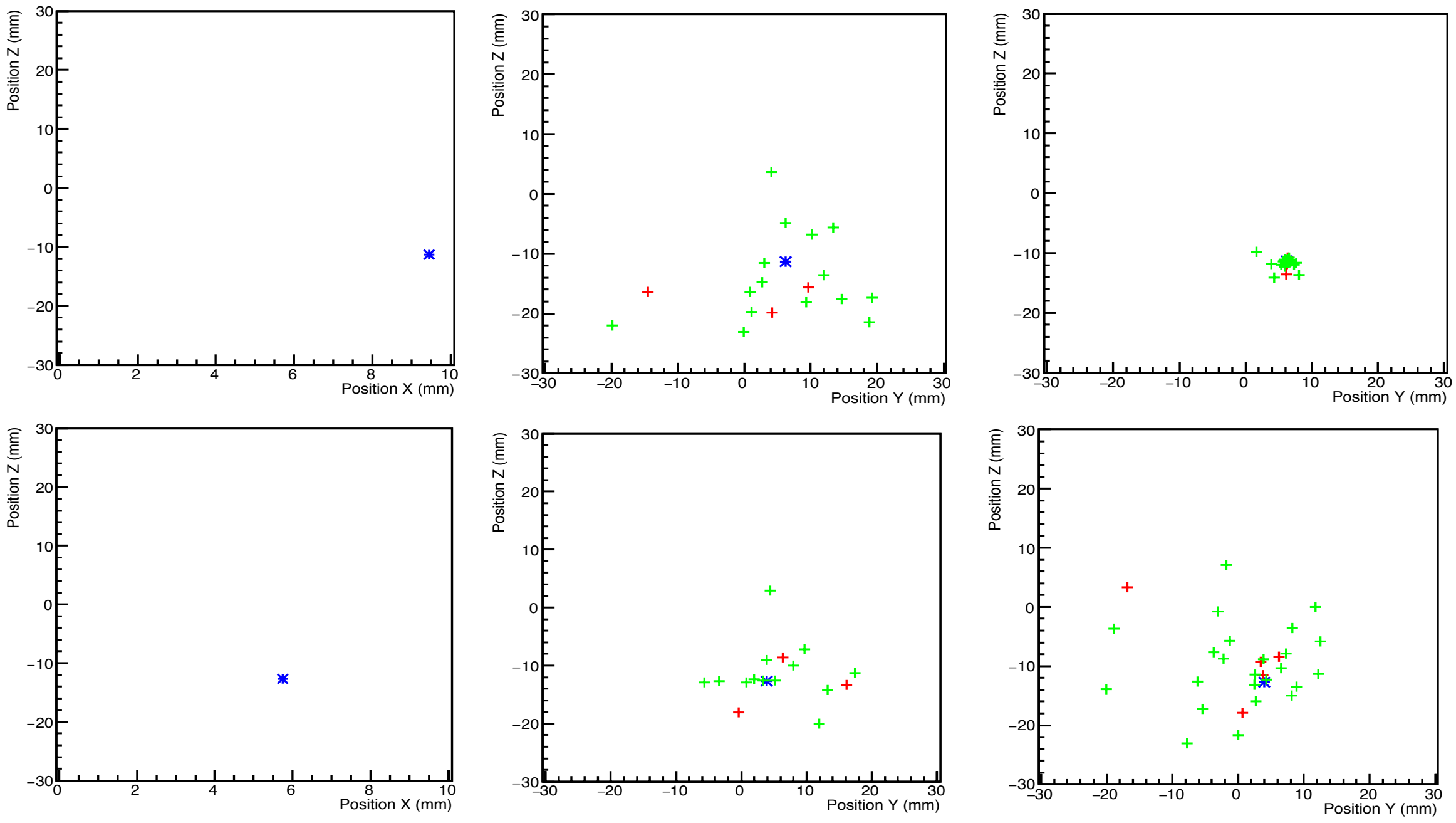

**Figure 7:** Simulated data for a 10 mm thick "scintronic" crystal instrumented on two sides by a bialkali photocathode. 511 keV gamma rays impinge on the detector from the left and propagate along the X axis. The upper and lower rows correspond to the photoelectric interaction of gamma rays close to the surface of the crystal and at mid-crystal depth, respectively. Notice the coordinates axis of each of the figures.
**(left)** Blue crosses denote gamma ray interactions within the crystal. **(centre and right)** show photoelectrons produced on the two photocathodes. Red and green crosses mark photoelectrons produced at X = 0 mm (center) and X = 10 mm (right) during the first 500 ps and the following 20 ns, respectively.

Figure 7 shows the computed photoelectron maps produced at the photocathode layers, which will be referred to as *hit-maps*. Typically, for a photoelectric interaction of a 511 keV gamma ray within $PbWO_4$, according to Geant4, under the assumptions stated above ($PbWO_4$:Y optical transparency stopping at 350 nm, refraction index varying from 2.3 to 2.1 for optical wavelength), 20 Cherenkov photons and 150 scintillation photons are produced. Monte Carlo simulation predicts an average of 20 photoelectrons produced by each photocathode. When an interaction takes place close to the surface of the detector, the hit-map corresponding to this face is very dense. In contrast, the hit-map corresponding to the opposite face of the crystal is spread on a large area. When the interaction takes place at mid-crystal depth, the apparent density of both hit-maps look intermediate.

Quantifying the properties of these hit-maps (barycenter, dispersion of hits around the barycenter, photon detection times, etc.) can then be used to assess the spatial and temporal properties of gamma ray interactions within the crystal (hidden physical information of interest).

## i. Preliminary event reconstruction

In this preliminary study, we have quantified on the hit-maps the number of photoelectrons, the 2D barycenter of the photoelectron positions, the 2D spread of the photoelectron positions around the barycenter, the time of the first photoelectron and the time distribution of prompt (within the first 500 ps) photoelectrons.

From the gamma ray interaction data, we have extracted the 3D coordinates and the time of the first gamma ray interaction, the total energy deposited in the crystal, and when there was more than one interaction, the barycenter of interaction positions weighted by the energy deposited at each interaction point.

Because of the low scintillation yield of $PbWO_4$, the number of photoelectrons provide a measurement of the energy deposited in the crystal of modest accuracy (~25 % FWHM), but compatible with a use in a PET scanner.

Classical (*i.e.* moment based, gaussian statistics) estimators (*e.g.* mean, variance) applied to hit-maps turned not to correlate accurately with the 3D coordinates of the gamma ray interactions. This is because, as it can be guessed from Figure 7, photoelectron distributions display long tails that impact strongly the computation of statistical moments. More "robust" statistical estimators have then to be used, the simplest being the medians of the distributions. Indeed, a barycenter defined as the median of the photoelectron coordinates along the Y and Z axis shows significantly improved correlations with the gamma interaction position. We have used:

$$Y_{bary} = \text{median}\{ Y_{PEi} \},\ Z_{bary} = \text{median}\{ Z_{PEi} \}$$

This is also true when quantifying the spread of hit-maps, for which we have used:

$$Y_{Spr} = \text{median}\{ |Y_{PEi} - Y_{bary}| \},\ Z_{Spr} = \text{median}\{ |Z_{PEi} - Z_{bary}| \}$$

From these values, we define the radial spread as:

$$RadSpr = (Y_{Spr}^2 + Z_{Spr}^2)^{1/2}$$

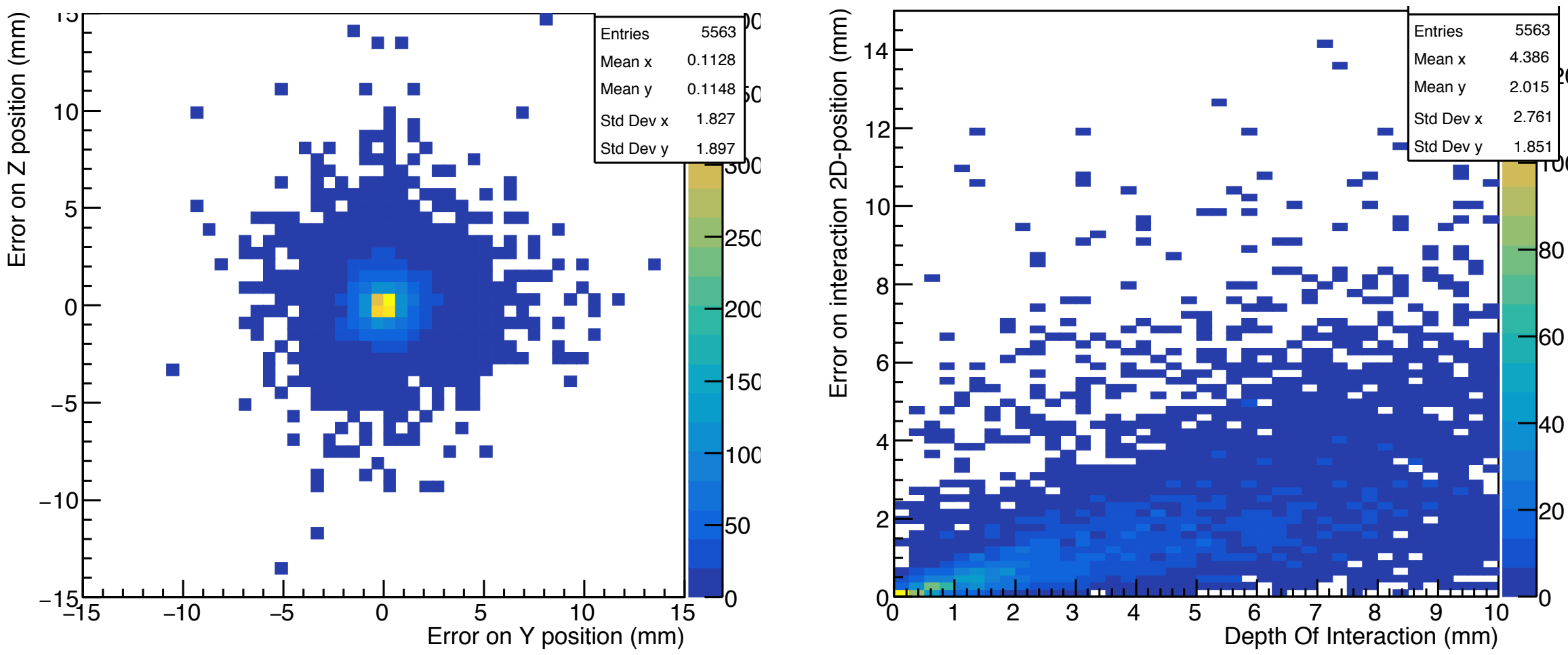

**Figure 8:** (Left) 2D histogram of the difference of computed Y and Z coordinates of the PE barycenter and the gamma ray position, using robust statistics. (Right) 2D histogram of $Dist_{baryPE}$ (see text for the definition) versus DOI . In these figures, we have used data from the front photocathode only. No selections were applied on the energy deposited in the $PbWO_4$ crystal.

## ii. Computed performance

In order to study the foreseen accuracy of a full (two faces instrumented) detector, we have assumed a $60 \times 60 \times 10$ mm$^3$ $PbWO_4$ crystal. We have simulated 511 keV gamma rays impinging orthogonally on the front face of the crystal at positions uniformly distributed within a centered square of $30 \times 30$ mm$^2$. This allows a preliminary study of the detector performances, without introducing the additional complexity of border effects. We have then computed the 2D (Y-Z coordinates) distance between the photoelectron barycenter and the gamma ray interaction position $Dist_{baryPE}$.

$$Dist_{baryPE} = ( (Y_{bary}-Y_{gamma})^2 + (Z_{bary}-Z_{gamma})^2 )^{1/2}$$

Figure 8 quantifies the correlation of $Dist_{baryPE}$ to the gamma ray interaction position, *using data from the front photocathode (X=0) only*. As it could be guessed from Figure 7, the larger the distance of the gamma ray interaction from the photocathode layer, the less accurate the reconstruction of the position.

The same study was made on the correlation between the measured radial spread, the time difference between the first photoelectrons detected by the photoelectric layers and the DOI of the gamma interaction within the crystal. Figure 9 shows some results.

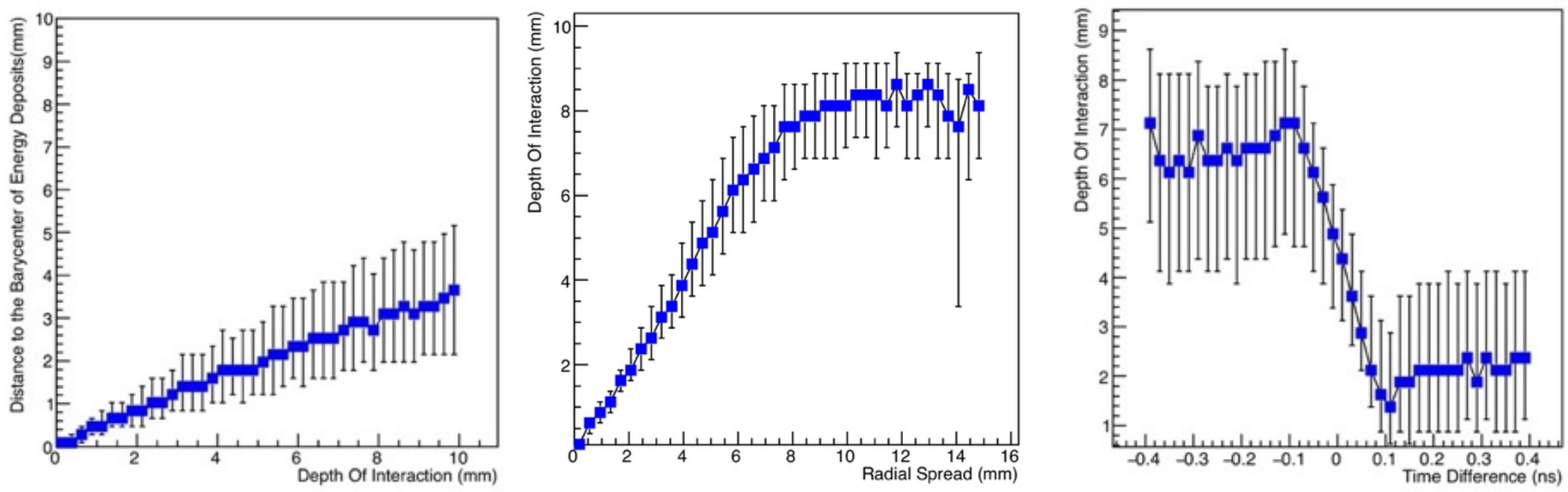

**Figure 9:** (Left) $Dist_{BaryPE}$ versus DOI. (Center) DOI as function of hit-map radial spread. (Right) DOI as function of time difference between the first photoelectron detected on the front and the rear photocathodes. For these three graphs, points show the mean values, bars on points correspond to the quartile value of the distribution statistics. No selections on energy deposited in the crystal were made.

From Figure 9, we notice that the gamma ray interaction parameters correlate with the detector observables in a complex way and that none of these observables are accurate enough to provide alone an accurate estimation of gamma ray interaction parameters. Left figure shows that the 2D (Y-Z coordinates) gamma ray interaction positioning accuracy depends a lot on the distance of the interaction to the photocathode. Center and left plots show the complex correlation between the gamma ray DOI (parameter to be reconstructed) and some of the detector observables.

Nevertheless, there are many observables into which the information is encoded. In order to extract more efficiently those physical parameters, we have chosen to use the resources of machine learning algorithms.

For a quick overview of available machine learning methods available, we have used the Toolkit for Multivariate Data Analysis (TMVA V4.2.0) library [48] as delivered within the ROOT modular scientific software toolkit package [49]. This package turned out to be very convenient: once the data have been preprocessed and formatted as a spreadsheet, we were able to select a collection of algorithms and to get a quick assessment of their efficiencies within few days. In this preliminary study, we have selected the KNN (k-Nearest Neighbour), FDA-GA (Functionnal Discriminant Analysis), MLP (Multi-Layered Perceptron) and Boosted Decision Tree (BDTG) methods. We have focused this work on the reconstruction of the 3D-coordinates, time and energy of the gamma ray interactions within the crystal.

We have assumed that a photocathode was deposited and instrumented directly on the front and rear faces of the crystal. Then we have fed the algorithms with the following statistical variables extracted from the simulated hit-maps:

- $Y_{bary}$, $Z_{bary}$, $Y_{Spr}$, $Z_{Spr}$
- the time of the first detected photoelectron
- the number of fast photoelectrons (*i.e.* detected within the first 500 ps)
- the number of slow photoelectrons (*i.e.* detected after 500 ps)

to which we have added the total number of photoelectrons detected in the event.

We have found that MLP and BDTG performed best with our simulated data giving very similar performances. MLP required a significant computation time, whereas BDTG procured a significantly faster training time, but its training algorithm happened to break with the library version we have used. It has shown also a tendency to do overtraining. Therefore, for the results which will follow, we have chosen to use a MLP based algorithm, *i.e.* a classical neural network structure.

In order to minimize the complexity of the work, we have decided to train a dedicated network for each parameter of the gamma ray interaction we were aiming at. This allows to simplify the network structure while keeping good performances. We have tried a handful of configurations, selecting the one that has provided fast training times and good performances. Hence, we have chosen a structure with two hidden layers and the hyperbolic tangent activation function, resulting in a $15 \times 5 \times 3 \times 1$ network structure. In this configuration on a recent laptop computer (thus without using GPU), we needed 22 s CPU time for the MLP algorithm training and 7.2 µs/event for event processing. Computer processing time for event reconstruction is thus not foreseen to be an issue with such an algorithm.

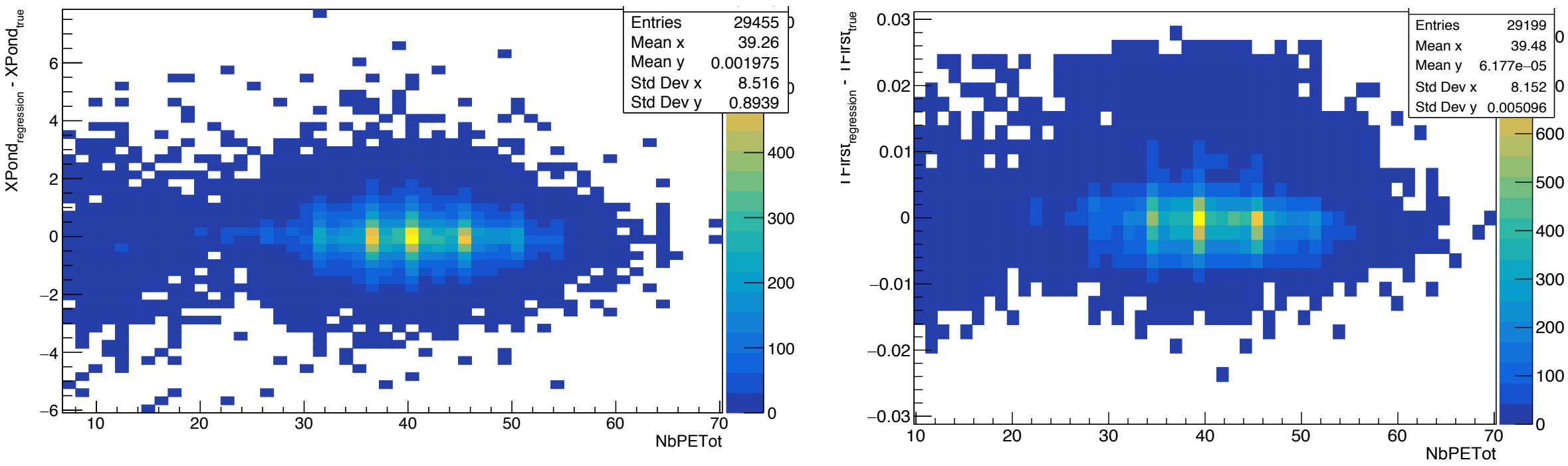

**Figure 10:** Evaluation of the performances of the MLP algorithms to reconstruct DOI and gamma ray interaction time within the $PbWO_4$ crystal from hit-map data. (Left**)** 2D histogram of the number of detected photoelectron versus the difference between reconstructed and simulated DOI (mm). (Right) 2D histogram of the number of detected photoelectron versus the difference between reconstructed and simulated time of the first gamma ray interaction (ns) within the crystal. No selection cuts were applied on simulated data.

Figure 10 shows 2D histograms of the reconstructed gamma ray interaction parameters versus the total number of photoelectrons produced in the photocathode. Both the histograms show very accurate performances, though simulated events include Compton and multiple interactions within the crystal.

**Table 3:** Average deviation (rms) computed on the DOI, lateral position, and time of the first gamma ray interaction with a MLP algorithm fed with robust statistics estimators. The first line gives the standard deviations computed from the complete statistics. The second line computes these standard deviations after removing the distribution tails larger than 2σ.

| Reconstructed parameter (rms) | DOI (mm) | Y coordinate (mm) | Time of the first gamma ray interaction (ps) | Energy deposit (keV) |
|---|---|---|---|---|
| Complete statistics | 0.91 | 1.57 | 5.5 | 27 |
| Truncated statistics | 0.69 | 1.28 | 4.2 | 7.3 |

Table 3 quantifies the results for all the gamma ray interaction parameters. The predicted accuracies are very motivating, though the simulated events include Compton and multiple interactions within the crystal. However, the energy deposit accuracy computation should be taken carefully. Indeed, the neural network has been trained assuming that all the impinging gamma rays had an energy of 511 keV. Thus, above the Compton edge, most energy deposits amount to 511 keV in the training sample and are identified as such, with zero reconstruction error in the test sample. Thus the computed energy resolution in this simulation is not representative of the energy resolution of the detector if it would have to be tested with random energy radioactive sources.

The reconstruction of the first gamma ray interaction time is very accurate and amounts to 13 ps FWHM, showing that the MLP algorithm is able to decorrelate appropriately time information from DOI. Nevertheless, our simulation was implementing neither the photodetector time transit spread, nor the readout electronics noise. Indeed, good quality MCP-PMT have single photoelectron time transit spread of 70 ps FWHM with extra 25 % events in a tail at the nanosecond time scale. Therefore, the above computed time resolution is quite optimistic and will be studied in details in future work. Forecasting its value will require detailed modelling of the full experimental readout chain. Nonetheless, this preliminary study has shown that machine learning algorithms have the potential to reconstruct accurately gamma ray interaction parameters when they are trained upon accurate Monte Carlo simulation.

# V. Discussion

In this paper, we propose a new concept of a "scintronic" PET detection module. We show that the main key technologies for its assembly have been demonstrated and we present a Monte Carlo study showing the strong potential of this "scintronic" detector design in terms of time resolution, efficiency and spatial resolution.

Further work will consist in assembling a detector with a relatively thin (~5 mm) $PbWO_4$ scintillating crystal used as the input window of a standard 2" squared MCP-MT. This design will allow us to validate the full "scintronic" detector assembling process, to learn how to optimize signal readout both from the hardware and software points, and hopefully to achieve competitive detector performance in terms of time and spatial resolutions. For detectors thicker than 5 mm, time resolution and overall event reconstruction performance will significantly improve when both the front and the rear faces of the $PbWO_4$ crystal will be instrumented. However, we foresee that integrating a thick scintillating crystal in between two MCP-PMTs will be difficult and therefore, we are also considering to use either a SiPM matrix or a regular MCP-PMT assembled with a standard optical coupling for the second face.

# Acknowledgments

The authors would like to thank J.-Ph. Renault from the CEA Saclay-IRAMIS and S. Chatain from CEA Saclay-DES for many advices related to chemistry and for their measurements, which have driven our choice of the detector crystal. We want to thank Prof. H. Frisch from Chicago University for stimulating discussions and for providing prototypes of transmission line readout boards. We also wish to thank Prof. M. Korjik from the Belarusian State University for discussions and advices on $PbWO_4$ scintillating crystal growth technologies. Finally, we wish to thanks R. Trebossen, S. Jan and C. Comtat from CEA- SHFJ for guiding us in the world of medical physics, and the Referee for his detailed review of the article.

We are grateful for the support and seed funding from the CEA "Programme Exploratoire Bottom-Up" under grant No. 17P103-CLEAR-MIND and the French National Research Agency under grant No. ANR-19-CE19-0009-01.